\documentclass[preprint2]{aastex}
\usepackage{amsmath}
\shorttitle{SALT2mu}
\shortauthors{Marriner et al.}

\begin{document}

\title{A More General Model for the Intrinsic Scatter in Type Ia Supernova Distance Moduli}

\author{
John Marriner,\altaffilmark{1}
J.P. Bernstein,\altaffilmark{2}
Richard Kessler,\altaffilmark{3,4}
Hubert Lampeitl,\altaffilmark{5}
Ramon Miquel,\altaffilmark{6,7} 
Jennifer Mosher,\altaffilmark{8}
Robert C. Nichol,\altaffilmark{5}
Masao Sako, \altaffilmark{8}
Donald P. Schneider,\altaffilmark{9} and
Mathew Smith\altaffilmark{10}
}
\altaffiltext{1}{Center for Particle Astrophysics, Fermi National Accelerator Laboratory, P.O. Box 500, Batavia, IL 60510, USA; marriner@fnal.gov}
\altaffiltext{2}{Argonne National Laboratory, 9700 South Cass Avenue, Lemont, IL 60439, USA.}
\altaffiltext{3}{Department of Astronomy and Astrophysics, The University of Chicago, 5640 South Ellis Avenue, Chicago, IL 60637, USA.}
\altaffiltext{4}{Kavli Institute for Cosmological Physics, The University of Chicago, 5640 South Ellis Avenue, Chicago, IL 60637, USA.}
\altaffiltext{5}{Institute of Cosmology and Gravitation, Dennis Sciama Building, Burnaby Road, University of Portsmouth, Portsmouth, PO1 3FX, UK.}
\altaffiltext{6}{Instituci\'o Catalana de Recerca i Estudis Avan\c{c}ats, E-08010 Barcelona, Spain.}
\altaffiltext{7}{Institut de F\'{\i}sica d'Altes Energies, E-08193 Bellaterra (Barcelona), Spain.}
\altaffiltext{8}{Department of Physics and Astronomy, University of Pennsylvania, 209 South 33rd Street, Philadelphia, PA 19104, USA}
\altaffiltext{9}{Department of Astronomy and Astrophysics, The Pennsylvania State University, 525 Davey Laboratory, University Park, PA 16802, USA.}
\altaffiltext{10}{Astrophysics, Cosmology and Gravitation Centre (ACGC) 
University of Cape Town, Rondebosch, Cape Town, Republic of South Africa.}

\slugcomment{July 22, 2011}

\begin{abstract}
We describe a new formalism to fit the parameters $\alpha$ and $\beta$ that are used in the SALT2 model to determine the standard magnitudes of Type Ia supernovae.  The new formalism describes the intrinsic scatter in Type Ia supernovae by a covariance matrix in place of the single parameter normally used.  We have applied this formalism to the Sloan Digital Sky Survey Supernova Survey (SDSS-II) data and conclude that the data are best described by $\alpha=0.135_{-.017}^{+.033}$ and $\beta=3.19_{-0.24}^{+0.14}$, where the error is dominated by the uncertainty in the form of the intrinsic scatter matrix.   Our result depends on the introduction of a more general form for the intrinsic scatter of the distance moduli of Type Ia supernovae than is conventional, resulting in a larger value of $\beta$ and a larger uncertainty than the conventional approach.  Although this analysis results in a larger value of $\beta$ and a larger error, the SDSS data differ (at a 98\% confidence level)  with $\beta=4.1$, the value expected for extinction by the type of dust found in the Milky Way.  We have modeled the distribution of supernovae Ia in terms of their color and conclude that there is strong evidence that variation in color is a significant contributor to the scatter of supernovae Ia around their standard candle magnitude.   
\end{abstract}


\section{Introduction}
Type Ia supernovae (SN) have been extensively used as ``standard candles'' to measure astronomical distances.  However, they are often called ``standardizable candles'' to emphasize the fact that the light curve shape and color can be used to correct the apparent  magnitudes, resulting in a more accurate measurement of the distance.  The correlation between light curve shape and magnitude was demonstrated by \citet{Ph93}, who parameterized the light curve shape with the parameter ${\Delta}m_{15}$, the decay in $B$-band magnitude 15 days after the peak.   Different parameterizations of the light curve shape parameter have been used: $\Delta$ for MLCS2k2 \citep{Jh07}, $x_1$ for SALT2 \citep{Gu07,Gu10}, and $s_B$ for SiFTO \citep{Co08}, but the parameter is essentially determined empirically to produce the tightest correlation between light curve shape and distance.  This work will use the SuperNova ANAlysis (SNANA) software package \citep{K09a} implementation of the SALT2 model that uses $x_1$ as the light curve shape parameter. 

Type Ia SN color depends on the light curve shape parameter, but SN also exhibit variations in color for fixed light curve shape.  The source of color variation is not well understood, but it is widely suspected that variations in the explosion process, host galaxy dust, and the circumstellar environment all play a role.  The difference in measured color and the mean color for a given $x_1$ is denoted by the parameter $c$.  Regardless of the process that produces different colors, it may be expected that the apparent $B$-band magnitude ($m_B$) will show some dependence on color, and that an appropriate color correction should improve the precision of SN as standard candles.  These considerations have led to the widespread practice of fitting SN light curves  with a one parameter family of templates plus a color law to describe the color.  

The determination of color is straight-forward given multi-band photometric measurements and a color law, and the result is traditionally given as $E(B-V)$.  However, the amount of correction to apply to a given SN has been less clear.  The correction has been often expressed as an extinction correction, which is related to the color through the parameter $R_V=A_V/E(B-V)$, where $A_V$ is the extinction in $V$-band.  Many efforts to determine this correction have relied on the parameterization of Cardelli, Clayton, and Mathis \citep{CCM}, which we will call the CCM model, where the parameter $R_V$ changes both the shape of the extinction curve as a function of wavelength and the amount of extinction for a given color excess $E(B-V)$.  The simplest approach \citep{Jh07} is to fix $R_V=3.1$, the average value determined from the studies of extinction in the Milky Way \citep{Sn78}.  However, many studies of SN light curve data have indicated that smaller values of $R_V$ provide a better description of the SN data.  In the CCM model the single parameter $R_V$ can be measured by estimating $A_V$ or by comparing the relative extinction in different filters (by considering $E(B-V)-E(V-R)$, for example). 

The SALT2 model parameterizes the SN rest-frame spectrum $S(\lambda,p)$ as a function of wavelength ($\lambda$) and phase ($p$) according to
\begin{equation}
S(\lambda,p)=x_0 \, [S_0(\lambda,p)+x_1 S_1(\lambda,p)] \, e^{[-cCL(\lambda)]},\label{eqn:salt2}
\end{equation}
where the fitted parameters $x_0$, $x_1$, and $c$ are the overall scale, the light curve shape parameter, and the color respectively.  The functions $S_0$ and $S_1$ are fixed by a training sample of SN light curves and spectra.  The SALT2 light curve fit does not rely on the CCM parameterization, but uses a color law $CL(\lambda)$ determined empirically from SN data.  The SALT2 parameters are determined by fitting the measured light curve data for each SN to the synthetic magnitudes calculated from the redshifted spectrum given by Equation \eqref{eqn:salt2}.

The light curve parameters $x_0$, $x_1$, and $c$ specify a model SN spectrum, but an additional step is needed to use the light curve parameters to calculate a standard SN magnitude. One can calculate the apparent magnitude $m_B$, where $m_B$ is the apparent $B$-band synthesized from the rest-frame spectrum given by Equation \eqref{eqn:salt2}, and a standardized SN brightness $m_B^s$ is computed with parameters $\alpha$ and $\beta$: 
\begin{equation}
m_B^s=m_B + \alpha x_1 - \beta c.\label{eqn:basic}
\end{equation}
\citet{Gu07} describe a technique that simultaneously determines the cosmological parameters and the coefficients ($\alpha$ and $\beta$) that minimize the residuals on the Hubble diagram.  The SALT2 approach to correction for color is formally equivalent to the description in terms of $R_V$ and $E(B-V)$, except for the interpretation that the value of $\beta$ may result from some effect other than extinction by host galaxy dust.  Regardless of the interpretation, $\beta$, which gives the magnitude correction for $B$-band, and $R_V$, which gives the magnitude correction for $V$-band, may be compared via the relation $\beta=R_B=R_V+1$.    

In this work, we present a new method for fitting the parameters $\alpha$ and $\beta$ from the data.  The method has been implemented in the program SALT2mu, which has been added to the SNANA software package.  The motivations for a new program are:  
\begin{itemize}
\item The fit for $\alpha$ and $\beta$ is decoupled from the cosmology.  If the mean values of $x_1$ and $c$ are a function of redshift (a magnitude limited survey will have such selection effects), the average residual depends on both the cosmology and the $\alpha$ and $\beta$ parameters.  We prefer to determine the $\alpha$ and $\beta$ solely by minimizing the rms of the distribution and to let the mean of the distribution determine the cosmology.
\item A new description of intrinsic scatter is used as discussed below.
\item The SALT2mu output can be used interchangeably with different cosmology fitters since SALT2mu produces distance moduli separately from the cosmology fit.  This feature allows a direct comparison of SN distance moduli without reference to a particular cosmology.
\item The sample used to determine $\alpha$ and $\beta$ can be different from the sample used to determine the cosmology.

\end{itemize}

A significant additional benefit to writing a new program is that SALT2mu reads files from the SALT2 light curve fits and outputs files that can be read by any of the SNANA cosmology fitters.  The fitting model resides in a single function making it easy to modify the $\chi^2$ function by adding new parameters, for example.  

The inclusion of a more general description of the intrinsic scatter of the fitted SALT2 parameters is a major feature, as is the determination of $\alpha$ and $\beta$ in a way that is independent of the cosmology.   The details of the incorporation of the intrinsic scatter are given in the mathematical description (\S\ref{sec:math}) below, and most of the paper is devoted to the issues associated with the more general description of the intrinsic scatter.  The program is then applied to the Sloan Digital Sky Survey supernova (SDSS-II SN) data, which is used to illustrate some of the features of the program and issues with its use.

The SDSS-II SN data \citep{Fr08} were obtained during 2004-2007 as part of extension of the original SDSS \citep{Yo00}.  The SDSS telescope \citep{Gu06} and imaging camera \citep{Gu98} produce photometric measurements in each of the 5 SDSS filters \citep{Fu96} spanning the range of 350 to 1000 nm.  The most useful filters for observing SDSS SN, however, are \textit{g}, \textit{r}, and \textit{i} because the SN are too faint in \textit{u} and \textit{z} to be well measured by the SDSS SN survey except at low redshifts.  Further details of the survey, data processing and selections are given in \S\ref{sec:SDSS}.

\section{Mathematical Description}\label{sec:math}

The standardized 10 pc supernova magnitude is determined by the expression
\begin{equation}
M_0^{(B)}=m_B-\mu(z)+\alpha(z) x_1 - \beta(z) c,\label{eqn:stBmag}
\end{equation}
where $\mu(z)$ is the distance modulus, and $M_0^{(B)}$ is the standardized SN magnitude for $B$-band.  We have generalized the unknown parameters $\alpha$ and $\beta$ to functions of redshift to accommodate possible effects of evolution of supernovae or their environment.  These functions  are not known from first principles, but must be determined from the data.  We use a nearly equivalent formulation
\begin{equation}
M_0(z)=m_x-\mu(z)+\alpha(z) x_1 - \beta(z) c,\label{eqn:stmag}
\end{equation}
where
\begin{equation}
m_x = -2.5\log_{10}(x_0).\label{eqn:mx}
\end{equation}
The definition shown in Equation \eqref{eqn:mx} has the advantage that the magnitude $m_x$ does not depend on the shape of the $B$-band filter.  It differs from the $B$-band standard magnitude by about 10 magnitudes.  The functions $M_0(z)$, $\alpha(z)$ and $\beta(z)$ are to be found by the minimization of the expression for $\chi^2$
\begin{multline}
\chi^2 = \displaystyle\sum_{n=1}^N 
[m_{xn}-\mu(z_n)+\alpha(z_n)x_{1n}-\beta(z_n)c_{n} \\
-M_0(z_n)]^2/(\sigma_n^2+\sigma_{\rm{int}})^2 
\label{eqn:s2mchi2}
\end{multline}
where $m_{xn}$, $x_{1n}$ and $c_{n}$ are the fitted SALT2 parameters for the $n^{\rm{th}}$ SN, and $N$ is the number of SN.  The quantity $\sigma_n$ is the error on the quantity in the numerator due to measurement uncertainty for the $n^{\textrm{th}}$ SN.  The quantity $\sigma_{\rm{int}}$ is discussed below and its form is displayed in Equation \eqref{eqn:sigmai}.  The function $M_0(z)$ is, by construction, degenerate with cosmological information contained in the distance modulus $\mu(z)$.  The introduction of a nuisance function $M_0(z)$ allows our fit to minimize the scatter in the Hubble diagram while being insensitive to the cosmology.  In order to reduce this function to a few parameters we assume that $M_0(z)$ is an arbitrary constant in each of several redshift bins centered at $z=z_b$, and we use the exact expression for the distance modulus $\mu(z)$ for some assumed cosmology.  Effectively we are using an assumed cosmology to extrapolate the residual to the redshift at the center of the bin and minimizing the Hubble diagram scatter at those reference points.  Since our method introduces several new parameters, the statistical accuracy of the fit is reduced by the additional parameters.  And since our method relies on the minimization of the scatter of distance moduli in a single bin, it requires a bin population of at least two SN---preferably many more.  These considerations are of no consequence for the SDSS sample, but may be of some concern in applying our method to samples with sparse data or unconventional cosmologies.  We show in \S\ref{sec:sim} that the fit results are insensitive to the choice of cosmological parameters if the number of redshift bins is sufficiently large.

Although this formalism allows fitting an arbitrary redshift dependence\footnote{SALT2mu includes the possibility of fitting a linear dependence for $\alpha$ and $\beta$.} for $\alpha$ and $\beta$, we will assume that $\alpha$ and $\beta$ are constants, independent of redshift, in this paper.  Recently, a number of papers \citep{Ke10,La10,Su10} have found that the value of $M_0$ shows a dependence on host galaxy properties.  Other papers have found correlations between line velocities in SN spectra \citep{Wa09,Fo11} and their standardized magnitudes and other correlations with spectral features \citep{No11}.  Those aspects will not be explored in this paper,  but it is straightforward to generalize Equation \eqref{eqn:stmag} by adding a new term $\gamma y_n$ to the numerator of Equation \eqref{eqn:s2mchi2}, where $y_n$ is some measured property of the $n^{\textrm{th}}$ SN and $\gamma$ is a parameter to be determined by the minimization.  In addition, the measurement uncertainty and intrinsic scatter in $y_n$ should be incorporated into $\sigma_n$ and $\sigma_{\rm{int}}$, respectively, following the formalism described below.

When the SALT2 light curve parameters are used according to Equation \eqref{eqn:stmag} to construct a SN Hubble diagram, the dispersion exceeds that expected from the light curve fit parameter errors.   A simple way to parameterize the Hubble diagram scatter is to assert that the distribution of light curve parameters can be described by an arbitrary population in the plane described by Equation \eqref{eqn:stmag} plus a Gaussian spread that is described by a covariance matrix, which we will refer to as the \textit{intrinsic scatter}.  The quanitity $\sigma_{\rm{int}}$ in the denominator of Equation \eqref{eqn:s2mchi2} describes the expected error in the numerator due to the intrinsic scatter.  We further assume that the covariance matrix is constant, independent of the SN light curve parameters.  These considerations result in an expression for the intrinsic scatter
\begin{multline}
\sigma_{\rm{int}}^2 = \Sigma_{00} + \alpha(z)^2\Sigma_{11}+\beta(z)^2 \Sigma_{cc}
+2\alpha(z)\Sigma_{01} \\
-2\beta(z)\Sigma_{0c}-2\alpha(z)\beta(z)\Sigma_{1c},\label{eqn:sigmai}
\end{multline}
where \textbf{$\Sigma$} is the intrinsic covariance matrix of the parameters and $\Sigma_{00}$, $\Sigma_{11}$, and $\Sigma_{cc}$ are the diagonal elements that correspond to the SALT2 parameters $m_x$, $x_1$, and $c$.  Our introduction of a covariance matrix to describe the intrinsic scatter differs from the conventional approach where the intrinsic scatter is assumed to be a single number.  Equation \eqref{eqn:sigmai} shows that the conventional approach is equivalent to our model if $\Sigma_{00}$ is the only non-zero term, leading to the conclusion that the conventional approach applies the intrinsic scatter to $m_x$ but not to $x_1$ or $c$.  We introduce the more general approach because the additional terms have not been excluded on either experimental or theoretical grounds and, in fact, there is good reason to believe that there is a significant scatter that can be attributed to the SALT2 color parameter.  The measurement error term ($\sigma_n^2$) in Equation \eqref{eqn:s2mchi2} is dominated by photometric errors and is calculated by the SALT2 light curve fit for each SN and is given by a covariance matrix that is of the same form as the intrinsic scatter covariance displayed in Equation \eqref{eqn:sigmai}.  The denominator in Equation \eqref{eqn:s2mchi2} is therefore the sum of a constant intrinsic scatter covariance matrix, which we have introduced above, and the measurement covariance, which has traditionally been included.  The fact that the measurement error and the intrinsic error are both of the form shown in Equation \eqref{eqn:sigmai} means that each of the SALT2 parameters has both an intrinsic scatter and a measurement error that are taken into account in computing $\chi^2$.  The formulation of the minimization problem defined by Equations  \eqref{eqn:s2mchi2} and \eqref{eqn:sigmai} is an extension of the formula given in \citet{Pr07} for the case of fitting a line to two variables with known errors.  A more extensive treatment of the linear regression problem from an astronomical perspective has been given, for example, by \citet{Ke07}. 

 After the parameters $\alpha$ and $\beta$ are determined by the minimization of $\chi^2$ the distance moduli are calculated according to
\begin{equation}
\mu_n = m_{xn}+\alpha(z_n) x_{1n}-\beta(z_n) c_{n}-M_0(z_b)\label{eqn:mu}
\end{equation}
These distance moduli can then be used for input to cosmological fits.

\section{Application to the SDSS data}\label{sec:SDSS}

We now apply this method to the SDSS 4-year sample of 529 spectroscopically confirmed SN.  The SDSS SN survey is a so-called ``rolling search,'' where a portion of the sky is repeatedly scanned to discover new SN and to measure the light curves of ones already discovered.  The survey measured an equatorial stripe about 2.5 degrees wide in declination between a right ascension of 20 h and 04 h.  Full coverage of the stripe can be obtained in two nights, but the average cadence was about four nights because of inclement weather and interference from moonlight.   The data were obtained in the Fall of 2005, 2006, and 2007 and a smaller amount of data was taken in a test run in 2004.  The survey is sensitive to SN beyond a redshift of 0.4, but beyond a redshift of 0.2 the completeness and the ability to obtain accurate photometry deteriorates.  The SDSS camera images were processed by the SDSS imaging software and SN were identified via a frame subtraction technique.  More details and references can be found in \citet{Fr08}. The candidate selection and spectroscopic identification has been described by \citet{Sa08}.  The supernova photometry used in this analysis was obtained using the scene modeling technique of \citet{Ho08}, where the sequence of SN observations is modeled as a variable point source and a host galaxy background that is constant in time.

We select a sample of well-measured light curves for this analysis based on the criteria used by  \citet{K09b}.  We require
\begin{itemize}
\item Spectroscopically confirmed as a Type Ia SN using techniques similar to those used for the first year sample \citep{Zh08}
\item Redshift in the range $0.02<z<0.42$
\item A minimum number of measurements without any requirement on the measurement error and a minimum number of well-measured points, where well-measured means an error of 20\% or less (\textit{i.e.}, $S/N>5$)
\begin{itemize}
\item At least 5 well-measured points
\item At least 2 well-measured points in different filters
\item At least 1 point measured before peak light in $B$-band
\item At least 1 point measured later than 9.5 days in the rest frame after peak light in $B$-band
\end{itemize}
\item An acceptable fit to the SALT2 model (confidence level greater than 0.001)
\end{itemize}
These selections result in a total sample of 343 SNe with negligible contamination from sources that are not Type Ia SN.  The redshift range spans 0.036 to 0.419, with a mean redshift of 0.22.  The SDSS sample thus defined does not have uniform selection criteria as a function of redshift and, therefore, there is a significant change in the population as a function of redshift.  In particular, the spectroscopically-selected sample of SDSS SN is significantly biased towards bluer, brighter SN as the redshift increases; the selection bias is discussed in detail in \citet{K09b}.  To the extent that the data are described by the model of Equation \eqref{eqn:stmag} with an accurate parameterization of $\alpha(z)$ and $\beta(z)$ and the SALT2 fit parameters are unbiased, we expect to be insensitive to having a biased population.  
 
An application of SALT2mu to the SDSS data using four redshift bins yields the results shown in Table \ref{tab:meas}, which shows the fit $\chi^2$ for 337 degrees of freedom (343 SN), the fitted values of $\alpha$ and $\beta$, and their errors $\sigma_\alpha$ and $\sigma_\beta$.  In section \S{4} we show that four bins is adequate to decouple the SDSS data from the determination of the cosmological parameters.  The first row in Table \ref{tab:meas} shows the fit when the intrinsic scatter is assumed to be zero resulting in $\chi^2=1305$,  reproducing the well-known result that the scatter in the Hubble diagram is larger than what is expected from the measurement error alone.  Subsequent rows illustrate how the fitted parameters change with different assumptions on the nature of the intrinsic scatter. Each of the fit variables is assumed, in turn, to be totally responsible for the intrinsic scatter with the magnitude of the scatter being adjusted to achieve $\chi^2/dof\approx 1$.  In Table \ref{tab:meas} we have introduced new variables to indicate the size of the intrinsic scatter:  $\sigma_0=\sqrt{\Sigma_{00}}$, $\sigma_1=\sqrt{\Sigma_{11}}$, and $\sigma_c=\sqrt{\Sigma_{cc}}$.  From Table \ref{tab:meas} we see that the different assumptions on the nature of the intrinsic scatter result in rather different estimates of the $\alpha$ and $\beta$ parameters compared to the reported statistical error.  An assumed scatter in $m_x$ ($\sigma_0\neq 0$) decreases both $\alpha$ and $\beta$ compared to the case where the intrinsic scatter is taken to be zero.  An assumed scatter in $x_1$ ($\sigma_1\neq 0$) increases $\alpha$ by $15.7\sigma$ while decreasing $\beta$ by $1.4\sigma$ relative to the fit where the intrinsic scatter is assumed to be entirely due to $m_x$.  The opposite effect is seen when the scatter is assumed to be due to $c$ ($\sigma_c\neq 0$): $\beta$ increases by $7.1\sigma$ and $\alpha$ decreases by $3.1\sigma$.  Even more variation could be exhibited by a more general intrinsic scatter covariance matrix.  However, the variations in $\alpha$ and $\beta$ are not without limit.  The values are all the same sign and approximately the same magnitude.  Roughly speaking, we are changing our estimate of how much to correct the observed slope for the flattening caused by the scatter described by the intrinsic covariance matrix (Equation \eqref{eqn:sigmai}).
\begin{deluxetable}{lrrrrr}
\tabletypesize{\scriptsize}
\tablecaption{SALT2mu fits to the SDSS Data\label{tab:meas}}
\tablewidth{0pt}
\tablehead{
\colhead{Intrinsic Error} & \colhead{$\chi^2$} &  \colhead{$\alpha$} 
& \colhead{$\sigma_\alpha$} & \colhead{$\beta$} & \colhead{$\sigma_\beta$} 
}
\startdata
$\sigma_0=\sigma_1=\sigma_c=0 $              & 1305 & 0.1891 & 0.0064 & 2.820 & 0.050  \\
$\sigma_0=0.14$,$\sigma_1=\sigma_c=0 $  &  350  & 0.1465 & 0.0098 & 2.650 & 0.078  \\
$\sigma_1=0.70$,$\sigma_0=\sigma_c=0 $  &  330  & 0.3002 & 0.0200 & 2.405 & 0.116  \\
$\sigma_c=0.05$,$\sigma_0=\sigma_1=0 $  &  333  & 0.1330 & 0.0110 & 3.208 & 0.102  \\
\enddata
\tablecomments{The SALT2mu program is applied to the 343 selected SDSS SN.  The  values of $\alpha$ and $\beta$ and their calculated errors ($\sigma_\alpha$ and $\sigma_\beta$) are shown.  There are 6 fit parameters ($\alpha$, $\beta$, and the 4 $M_0$ values for the 4 redshift bins) so the number of degrees of freedom is 337.  The variable $\sigma_0$, $\sigma_1$, and $\sigma_c$ give the assumed rms scatter in the SALT2 parameters $m_x$, $x_1$, and $c$, respectively.  The first row in the table shows the fit assuming an intrinsic scatter of zero; the following 3 rows assume that the intrinsic scatter results, in turn, from a single SALT2 fit parameter. }
\end{deluxetable}

\begin{figure}
\epsscale{1.0}
\plotone{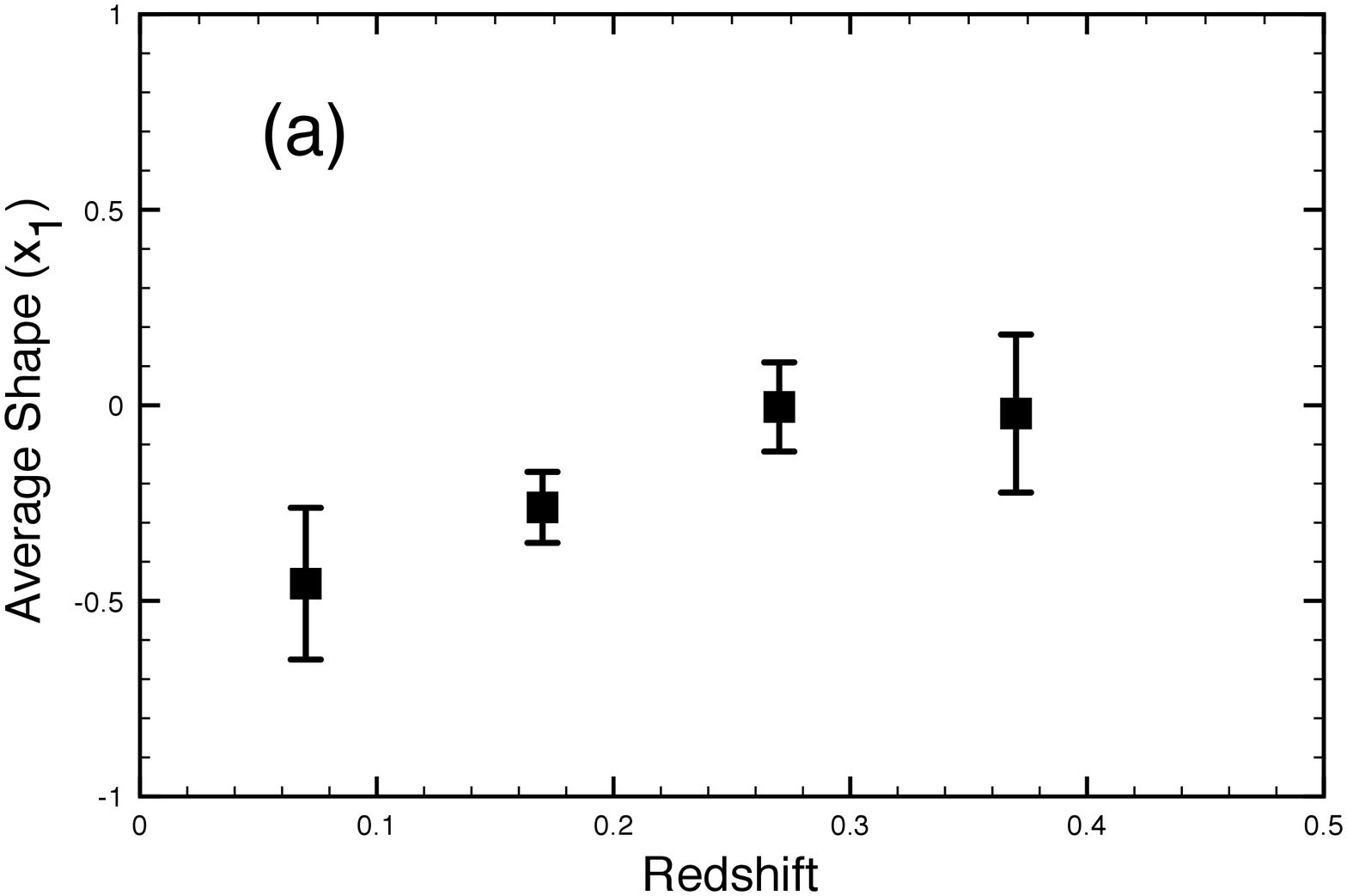}
\plotone{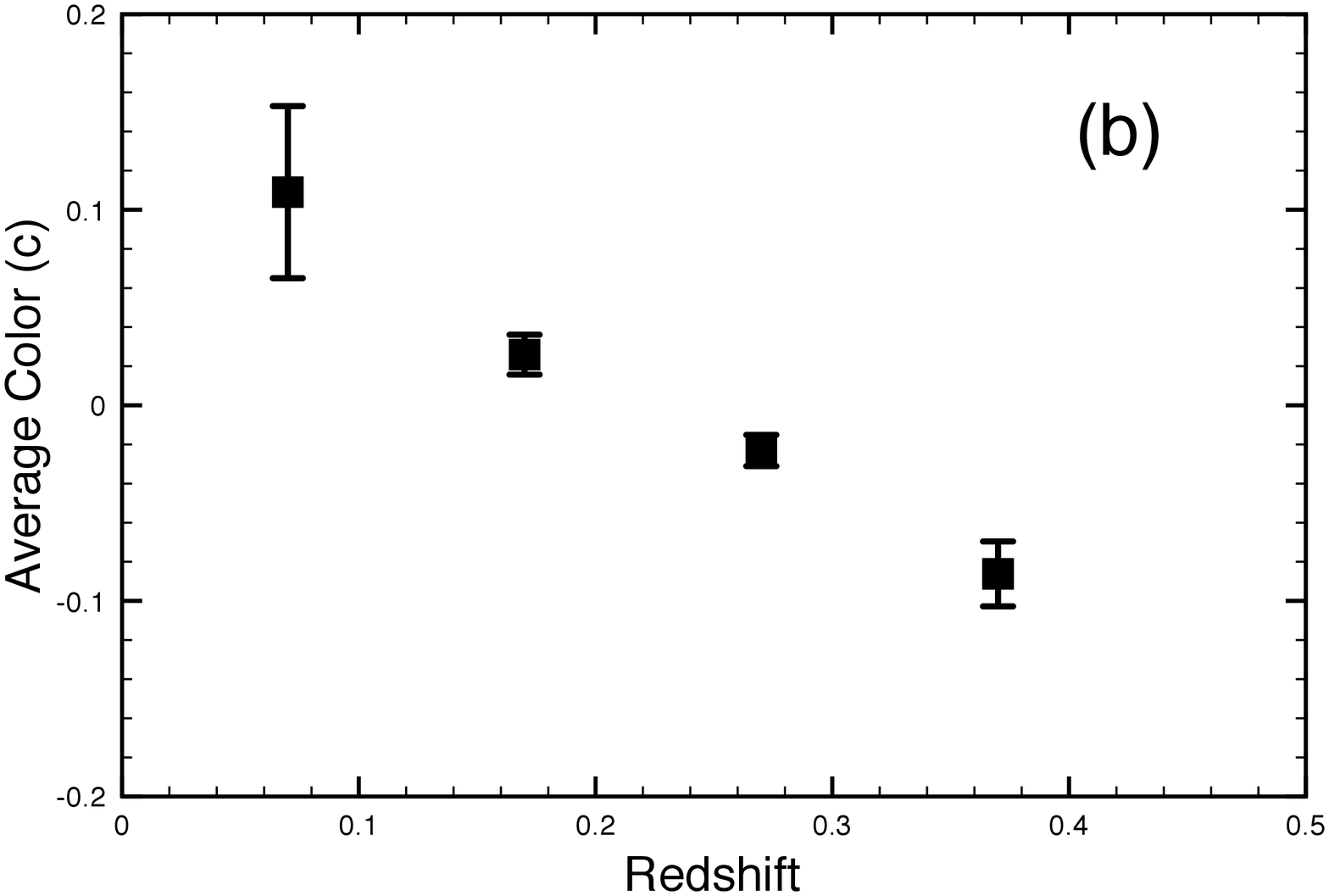}
\caption{The average value of (a) SALT2 shape ($x_1$) and (b) color ($c$) as a function of redshift for the SDSS sample of 343 SN.  The general trends of the data can be explained by selection effects.\label{fig:ave}}
\end{figure}

If we are only interested in SN as distance indicators, $\alpha$ and $\beta$ are nuisance parameters of no particular interest.  The average SN brightness at a given redshift will be given by the mean of Equation \eqref{eqn:stmag}:  if the mean values of $x_1$ and $c$ do not evolve with redshift, then different values of $\alpha$ and $\beta$ will result in a redshift-independent offset that is not significant in determining the cosmology.  The mean values of $x_1$ and $c$ for the SDSS data are plotted in Figure \ref{fig:ave} showing, however, a substantial increase in $x_1$ and a decrease in $c$ with redshift.  These trends are primarily caused by the spectroscopic-selection bias towards brighter SN Ia.  The different values listed in Table \ref{tab:meas} would result in differences of $\sim0.1$ in the average values of $\mu(z)$, a rather large error for cosmological studies.  In practice, the effect can be mitigated by limiting the sample to lower redshifts, where the selection efficiency is higher, correcting for biases, and combining high and low redshift experiments that have different efficiencies.  Nonetheless, the possibility of an uncorrected bias is a concern.

Our results can be compared to the results of \citet{Am10} who combined data from several sources to obtain a value of $\beta=2.51\pm0.07$ with an analysis most similar to our case with $\sigma_0=0.14$.  \citet{Gu07} found $\beta=1.77\pm0.16$ for the first year data from the SNLS and \citet{K09b} found $\beta=2.65\pm0.08$ for the first year of SDSS data.   \citet{Am10} also quote numbers for their four largest samples:  $\beta=2.38\pm0.14$ (SDSS),  $2.73\pm0.13$ (CfA), $2.50\pm0.17$ (ESSENCE), and $1.72\pm0.18$ (SNLS).  More recently \citet{Co11} found $\beta=3.18\pm0.10$ for the SNLS 3-year sample using a chi-squared minimization technique.
The study of \citet{No08} found the variations in SN colors were most consistent with a CCM parameterization when $R_V=1.75\pm0.27$.  However, the \citet{No08} study differs significantly from this analysis in that their value of $R_V$ is based on their measurement of the wavelength dependence of the reddening using the Cardelli parameterization.  
\citet{K09b} also investigated SN color and found $R_V= 2.18\pm0.14(stat)\pm0.48(syst)$, but that result relies on modeling the color changes due to the magnitude limited sample.
All these results are broadly consistent in finding $\beta<4.1$, but they exhibit more scatter than one would expect from the quoted errors.

\section{Tests with simulated SN samples}\label{sec:sim}

We will return to the SDSS SN data in \S\ref{sec:err}, but it is useful to first examine the performance of SALT2mu on simulated SN data.  We first determine the number of data bins that are required to decouple our fits from the cosmology over a wide range of parameters and conclude that 4 bins are adequate for the SDSS data.  We then verify the formalism with a simplified simulation and show that we recover the input values of $\alpha$ and $\beta$ without bias.  We next look at a realistic simulation and show that the fitted $\alpha$ and $\beta$ dependence on the intrinsic covariance matrix $\Sigma$ follow the trends seen in the data.  We compute the intrinsic scatter covariance matrix from the simulation (which requires knowing the simulation parameters), and show that the input values of $\alpha$ and $\beta$ are recovered when the correct intrinsic scatter covariance matrix is used.

We generate the simulated SNe by using SNANA to generate SN light curve data according to the SALT2 model.  The SN are generated with a redshift distribution appropriate for the SDSS survey and with $\alpha=0.18$ and $\beta=3.4$.  Realistic cadence, measurement errors, and selection effects are applied to the simulated SN,  and they are processed with the same light curve fitting program and selection criteria as the SDSS SN, resulting in SALT2 fit parameters for the simulated SN.  In addition to the photometric measurement errors, the simulation applies \textit{color smearing} to model the intrinsic scatter.   Color smearing consists of random magnitude offsets that are generated for each SN and observed filter and added to each light curve point measured in that filter.  The magnitude of the smearing is adjusted to 0.08 in each observed filter to give approximately the same Hubble diagram scatter as is seen in the real data.  

\begin{figure}
\epsscale{1.0}
\plotone{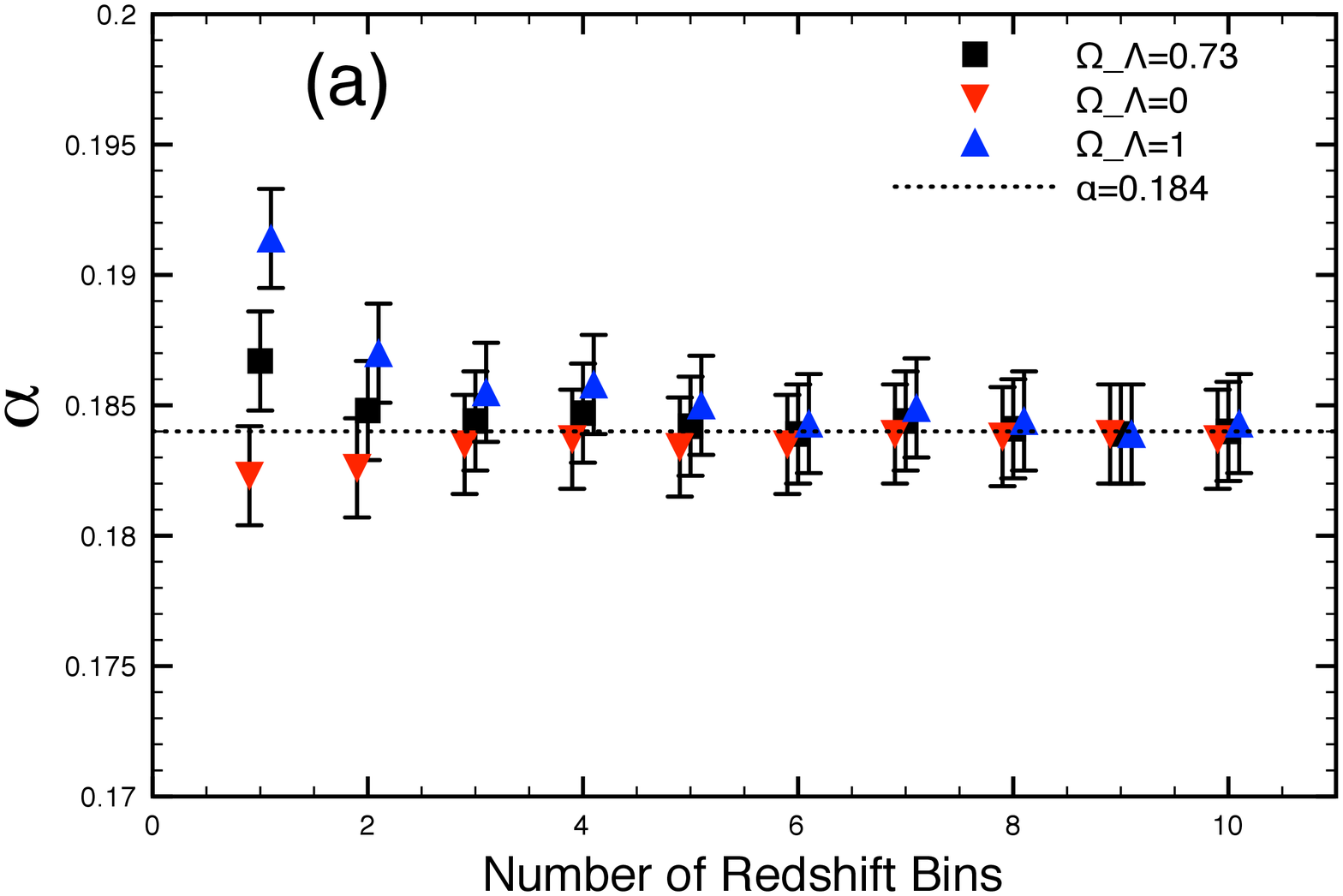}
\plotone{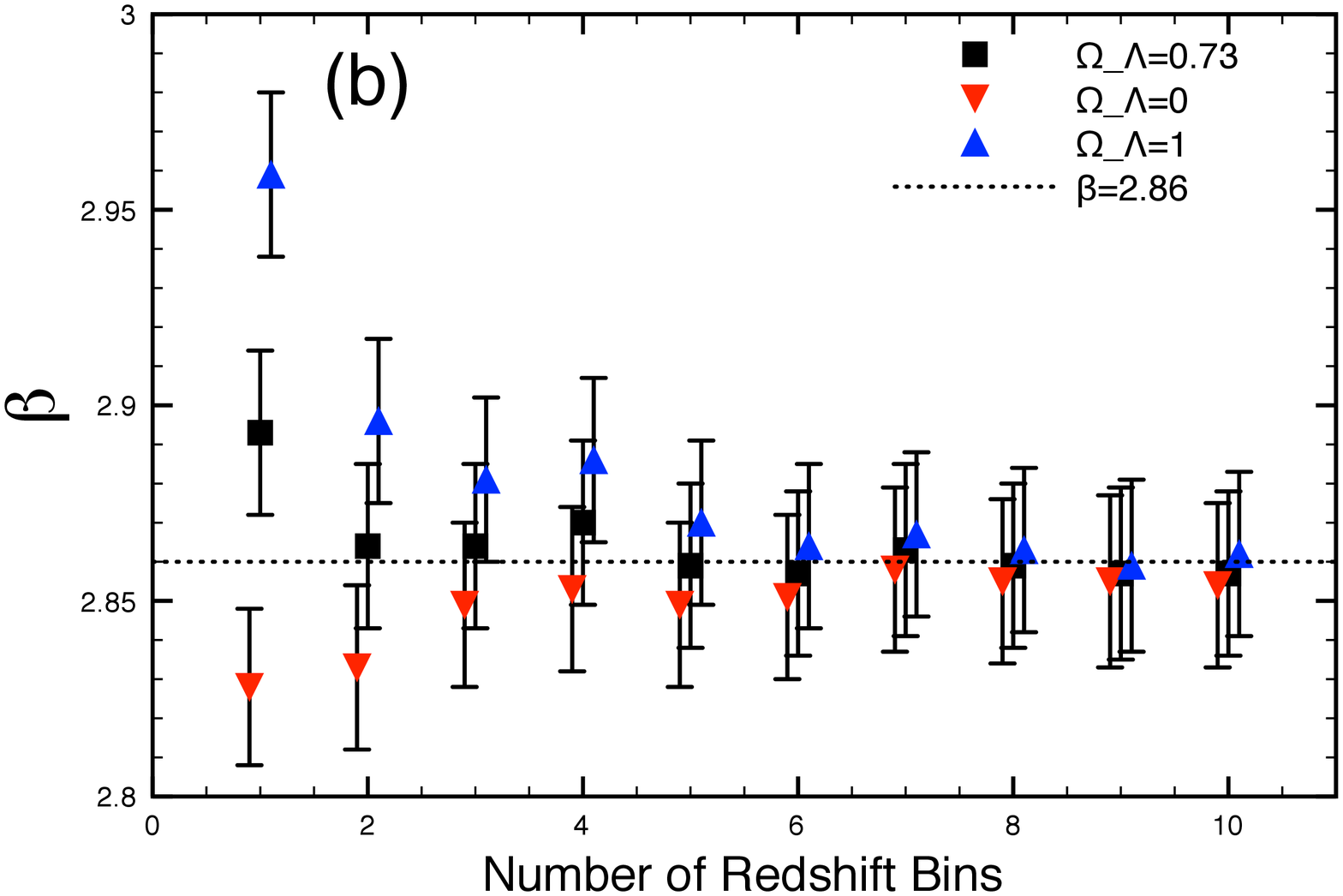}
\caption{The dependence of the fitted parameters $\alpha$ (a) and $\beta$ (b) is displayed as a function of the number of redshift bins for three assumptions about cosmological parameters and a particular choice of the intrinsic scatter matrix ($\sigma_0=0.14$, $\sigma_1=\sigma_c=0$).  As the number of bins increases, the fits converge to a common value $\alpha=0.184$ and $\beta=2.86$ shown as dotted lines on the respective figures.  \label{fig:ab}}
\end{figure}

The first test is a relatively trivial one to verify that the fit results for $\alpha$ and $\beta$ are, indeed, independent of cosmology.  We expect that a large number of parameters $M_0(z_b)$, will decouple the determination of $\alpha$ and $\beta$ from the cosmological parameters, but wish to minimize the number of nuisance parameters that are introduced.  We illustrate the fit dependence on the number of redshift bins in Figure \ref{fig:ab} using the simulation of the SDSS dataset, which has over ten times the number of SN (4417) as the SDSS data.  Figure \ref{fig:ab} shows the fit results for $\alpha$ and $\beta$ assuming three different cosmologies: $\Omega_M=0.27$, $\Omega_\Lambda=0.73$, $w=-1$ (red), $\Omega_M=0.27$, $\Omega_\Lambda=0.0$ (green), and $\Omega_M=0.27$, $\Omega_\Lambda=0.73$, $w=0$ (blue). The difference in the fitted parameters is noticeable when only 1 bin is used, but becomes negligible as the number of bins becomes greater than about 5.  We conclude that using 4 redshift bins is adequate for the SDSS data, given the lower statistical precision of the data and the established constraints on the cosmological parameters.  All results in this paper are based on dividing the SDSS data (real or simulated) into 4 equal sized bins spanning the redshift range $0.02<z<0.42$.   

While the fit results in Figure \ref{fig:ab} show that we have successfully decoupled the fit for $\alpha$ and $\beta$ from the cosmology, the dependence on the assumptions about the intrinsic scatter matrix remains.  For the fit results shown in Figure \ref{fig:ab} we took $\sigma_0=0.14$ and $\sigma_1=\sigma_c=0$, and this choice results in fitted values of $\alpha=0.184$ and $\beta=2.86$ which differ from those used to generate the simulated SNe, namely $\alpha=0.18$ and $\beta=3.4$.  The discrepancy is not surprising since our simulation includes significant color smearing that we have not taken into account.  We will determine more accurate parameters for the intrinsic scatter covariance below, but first we check the formalism with a simplified simulation.  We use the same simulated SN but replace the simulated values of $m_x$, $x_1$, and $c$ with the exact values that were used to generate the light curves but smeared according to the $3\times3$ error matrix including any assumed intrinsic scatter.  This procedure tests the self-consistency of the SALT2mu model and eliminates concerns about biases in the  SALT2 parameter fitting and inaccuracies in the calculation of the parameter errors.  The results of the fits are shown in Table \ref{tab:intsim} with the same four assumptions about the intrinsic scatter that were used in Table \ref{tab:meas}.    We have also tested the formalism when all the elements of the intrinsic scatter covariance matrix are non-zero; the last line in Table \ref{tab:intsim} is an example.  For each case we recover the input values that are consistent within the statistical error of about 1\%.   The $\chi^2/dof$ consistent with 1.  We do not observe any variation in the parameter values since the parameters of each SN are generated according to a Gaussian error distribution that is the combination of the measurement error and whatever intrinsic scatter was assumed.

\begin{deluxetable}{lrrrrr}
\tabletypesize{\scriptsize}
\tablecaption{SALT2mu fits to the simplified, simulated SDSS Data\label{tab:intsim}}
\tablewidth{0pt}
\tablehead{
\colhead{Intrinsic Scatter} & \colhead{$\chi^2$} &  \colhead{$\alpha$} 
& \colhead{$\sigma_\alpha$} & \colhead{$\beta$} & \colhead{$\sigma_\beta$} 
}
\startdata
$\sigma_0=\sigma_1=\sigma_c=0 $       &  4445  & 0.1802 & 0.0009 & 3.391 & 0.010  \\
$\sigma_0=0.14$                                &  4487  & 0.1774 & 0.0019 & 3.395 & 0.024  \\
$\sigma_1=0.70$                                &  4319  & 0.1815 & 0.0018 & 3.423 & 0.023  \\
$\sigma_c=0.05$                                &  4277  & 0.1818 & 0.0022 & 3.379 & 0.028  \\
Full covariance\tablenotemark{a}     &  4564  & 0.1811 & 0.0015 & 3.356 & 0.019  \\
\enddata
\tablenotetext{a}{The full covariance matrix used for this example was $\sigma_0=0.05$, $\sigma_1=0.35$, $\sigma_c=0.025$, $\xi_{01}=-0.2$, $\xi_{0c}=0.5$, and $\xi_{1c}=0.4$, where the $\xi$ are the correlation coefficients defined in the text following Equation \eqref{eqn:errtot}.
}
\tablecomments{In the SALT2mu simplified simulation, the SALT2 fit parameters ($m_x$, $x_1$, and $c$) are replaced with the true parameters plus an random error determined by the combination of the measurement and intrinsic scatter. The number of simulated SN is 4417 and the number of degrees of freedom for this fit is 4411.}
\end{deluxetable}

Using the parameters of the SALT2 fits to the simulated data provides a more complicated test since the color smearing model is different than applying an intrinsic scatter to the simulated SALT2 parameters used to generate the light curves.  We apply SALT2mu to the simulated SN using the same procedure used for the SDSS data, and display the results in Table \ref{tab:sim}.  The first line has the intrinsic scatter set to zero, and the succeeding lines have the intrinsic scatter that results in $\chi^2/dof\approx1$.  The general trends shown in Table \ref{tab:sim} are very similar to those seen in Table \ref{tab:meas}:  assuming an intrinsic scatter in $x_1$ and $c$ increases the fitted value of $\alpha$ and $\beta$, respectively.  The magnitudes of the intrinsic scatter required to achieve $\chi^2/dof\approx1$ are similar to those required by the SDSS data.  But we can also compare the results with the ``true'' values for the simulated data. While the assumption that the intrinsic scatter is entirely due to color errors (last line in Table \ref{tab:sim}) produces values of $\alpha$ and $\beta$ that are close to those used to generate the simulation ($\alpha$ differs by $2.6\sigma$), it is not a very accurate representation of the intrinsic scattering covariance matrix that describes the simulated SN as shown below.

\begin{deluxetable}{lrrrrr}
\tabletypesize{\scriptsize}
\tablecaption{SALT2mu fits to the simulated SDSS Data\label{tab:sim}}
\tablewidth{0pt}
\tablehead{
\colhead{Intrinsic Error} & \colhead{$\chi^2$} &  \colhead{$\alpha$} 
& \colhead{$\sigma_\alpha$} & \colhead{$\beta$} & \colhead{$\sigma_\beta$} 
}

\startdata
$\sigma_0=\sigma_1=\sigma_c=0 $       & 19500 & 0.2143 & 0.0011 & 3.104 & 0.011  \\
$\sigma_0=0.14$                                  &  4340 & 0.1846 & 0.0019 & 2.868 & 0.021  \\
$\sigma_1=0.65$                                 &  4375  & 0.2482 & 0.0025 & 2.882 & 0.026  \\
$\sigma_c=0.045$                               &  4343  & 0.1855 & 0.0021 & 3.370 & 0.025  \\
\enddata
\tablecomments{The first row in the table shows the fit assuming an intrinsic scatter of zero; the following 3 rows assume that the intrinsic scatter results, in turn, from a single SALT2 fit parameter.  The number of simulated SN is 4417 and the number of degrees of freedom for this fit is 4411. }
\end{deluxetable}

In principle, we know the smearing formulae applied in the simulation and can calculate the intrinsic scatter covariance matrix $\Sigma$.  But the calculation of the intrinsic smearing matrix is complicated because the error model used in the simulation smears individual light curve points in a way that is not easily expressed in terms of an intrinsic smearing matrix in $m_x$, $x_1$, and c.   However, it is relatively straightforward to calculate the error \textit{a posteriori}.  The full error matrix is determined by the variance of the simulated measurements of $m_x$, $x_1$, and $c$ relative to the true values (which, of course, are known for the simulated data).  The light curve fitter calculates covariance matrix of measurement errors, which, when added to the intrinsic scatter covariance matrix, should yield the total error matrix.  The intrinsic scatter covariance matrix can therefore be calculated for the simulated data by subtracting the observed variance from the prediction for the measurement error.  The total error covariance matrix calculated from the variance of the reconstructed SALT2 parameters from the simulated ones is

\begin{equation}
\begin{array}{lll}
\sigma_0= 0.0781, &  \xi_{01}=0.253,  & \xi_{0c}=0.612, \\
\sigma_1=0.351, & \xi_{1c}=0.183, &\textrm{and} \\
\sigma_c=0.0634. \label{eqn:errtot}
\end{array}
\end{equation}
where the various $\xi$ are the correlation coefficients ($\xi_{01}=\Sigma_{01}/(\sigma_0 \sigma_1)$, $\xi_{0c}=\Sigma_{0c}/(\sigma_0 \sigma_c)$, and $\xi_{1c}=\Sigma_{1c}/(\sigma_1 \sigma_c)$).  Using the same notation to represent the elements of the measurement error matrix we find
\begin{equation}
\begin{array}{lll}
\sigma_0= 0.0365, &  \xi_{01}=0.581,  & \xi_{0c}=0.694,  \\
\sigma_1=0.347, & \xi_{1c}=0.307, &\textrm{and} \\
\sigma_c=0.0286. \label{eqn:errmeas}
\end{array}
\end{equation}
Subtracting  \eqref{eqn:errmeas} from \eqref{eqn:errtot} gives an excess error, which we attribute to the intrinsic scatter.   The result is
 \begin{equation}
\begin{array}{lll}
\sigma_0= 0.0690, &  \xi_{01}=-0.119,  & \xi_{0c}=0.590,  \\
\sigma_1=0.0506 & \xi_{1c}=0.362, &\textrm{and}\\
\sigma_c=0.0566. \label{eqn:errint}
\end{array}
\end{equation}

We are now in position to apply the intrinsic scatter from Equation \eqref{eqn:errint} above to the fit from SALT2mu.  The result is shown in the first row in Table \ref{tab:fullerr}.  The result is consistent with the generated values ($\alpha=0.18$ and $\beta=3.4$).  The small discrepancy in $\beta$  ($2.6\sigma$) not a serious concern because it is quite small compared to the accuracy of the SDSS dataset.  However, we suspect that uncertainties in the calculation of the measurement errors, including treatment of non-linearities and a consistent handling of model errors, may contribute a systematic bias at this level of precision.  We have performed additional fits with other intrinsic scatter matrices.  The second row is a fit with all the non-diagonal elements of the intrinsic scatter matrix set to zero.  The fitted parameters are only slightly changed although there is a decrease in $\chi^2$.  The next row shows a fit that has been selected to give a larger change in $\chi^2$ by tuning the correlations for maximum effect.  The last row in Table \ref{tab:fullerr} displays a fit, which may be counter-intuitive:  the error in $x_1$ is decreased, but the $\chi^2$ increases.  These results illustrate how complicated the fit behavior can be for the five free parameters that live in the intrinsic scatter matrix.

 \begin{deluxetable}{lrrrrr}
\tabletypesize{\scriptsize}
\tablecaption{SALT2mu fits to the simulated SDSS data \break with non-zero, off-diagonal elements of the intrinsic scatter matrix \label{tab:fullerr}}
\tablewidth{0pt}
\tablehead{
\colhead{Intrinsic Error} & \colhead{$\chi^2$} &  \colhead{$\alpha$} 
& \colhead{$\sigma_\alpha$} & \colhead{$\beta$} & \colhead{$\sigma_\beta$} 
}
\startdata
As computed                                                          & 4161  & 0.1832 & 0.0022 & 3.469 & 0.027  \\
As computed but $\xi_{01}=\xi_{0c}=\xi_{1c}=0.0$                            &  2752  & 0.1822 & 0.0026 & 3.320 & 0.031  \\
As computed but $\xi_{01}=0.847,\xi_{0c}=-0.913,\xi_{1c}=-0.831$  &  1758  & 0.1820 & 0.0031 & 3.227 & 0.036  \\
As computed but $\sigma_1=0$                                         &  3130  & 0.1817 & 0.0024 & 3.400 & 0.030  \\
\enddata
\tablecomments{The first row shows the intrinsic scatter matrix computed as described in the text  and succeeding rows illustrate various changes to the intrinsic scatter matrix.  The simulation is the same as was used to produce Table \ref{tab:sim}.}
\end{deluxetable}

\section{Determination of the intrinsic scatter matrix}\label{sec:err}

We do not have a rigorous approach to estimating the intrinsic scatter matrix for actual SN data.  Since we don't know the underlying, unsmeared distribution or the actual form of the smearing, the observed distribution is difficult to interpret.  The scatter in the Hubble diagram sets the overall scale, but more information is needed to determine the individual elements of the intrinsic scatter matrix.

\begin{figure}
\epsscale{1.0}
\plotone{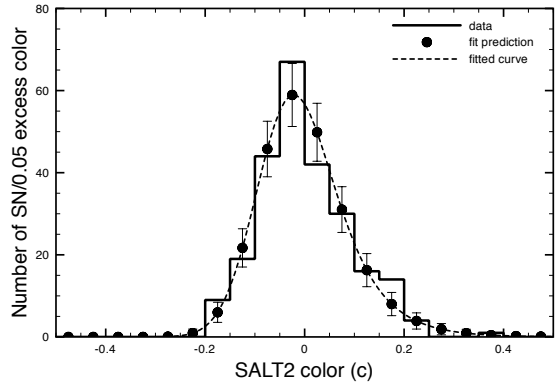}
\caption{The distribution in SALT2 color of 247 SDSS SN selected to have a calculated measurement error $\sigma_c<0.04$ is shown as a histogram, and the dashed curve is the best fit to the data.  Individual points on the curve are shown at the center of the histogram bins with errors corresponding to the expected fluctuations from Poisson statistics.\label{fig:color}}
\end{figure}

We can derive some information from the color distribution with the help of some plausible assumptions, which were first used by \citet{Jh07}.   We assume that the color distribution arises from random color smearing that we parameterize as a Gaussian of unknown width ($\sigma_c$) and host galaxy extinction that we characterize as an exponential distribution of unknown slope ($\tau$).  We do not expect to be sensitive to the exact shape assumed for galaxy extinction:  the important point is that it is one-sided (extinction can only make objects redder, not bluer) and that the most probable value of extinction is zero.  In addition, we do not know the position of the edge of the extinction distribution, so we must introduce an additional parameter ($c_0$) to be determined from the data.  Assuming that the intrinsic scatter and host galaxy extinction are independent, the convolution of the two processes results in the following distribution for $c$:
\begin{multline}
\rho (c)=A \int_{c-c_o}^\infty \exp\left(-\frac{c-c_o-c^\prime}{\tau}\right) \\
\times \exp\left(-\frac{{c^\prime}^2}{2 \sigma_c^2}\right) d c^\prime\label{eqn:cdist}
\end{multline}
In order to minimize the contribution of measurement errors to the width of the Gaussian component, we include only SN where the estimated measurement error for $c$ is less than 0.04, reducing our sample of 343 SN to 247.  Fitting Equation \eqref{eqn:cdist} to the data, we find $\sigma_c=0.061 \pm 0.007$, $\tau=0.076 \pm 0.010$, and $c_0=-0.073 \pm 0.009$.  The data and resulting fit are shown in Figure \ref{fig:color}.  As can be seen, the functional form of Equation \eqref{eqn:cdist} provides a good description of the data.  The fitted value for $\tau$ is substantially smaller than the value $\tau=0.334\pm0.088$ quoted by \citet{K09b}, but our result is an effective $\tau$ for the spectroscopically-selected sample with the additional requirement of a well-measured color and is therefore biased against large extinctions.  Measurement errors are calculated by the SALT2 light-curve fit and contribute, on average, $\sqrt{\langle \sigma_c^2 \rangle}=0.026$ to the width of the Gaussian component, resulting in an estimate of $0.055\pm0.007$ for the color smearing.    We do not expect our values of $c_0$ and $\sigma_c$ to be significantly affected by the selection bias, and we do not make a correction for the bias.  This result is comparable to the amount of color smearing ($\sigma_c=0.05$) that was required in Table \ref{tab:meas} to explain, by itself, the scatter in the Hubble diagram.

\begin{figure}
\epsscale{1.0}
\plotone{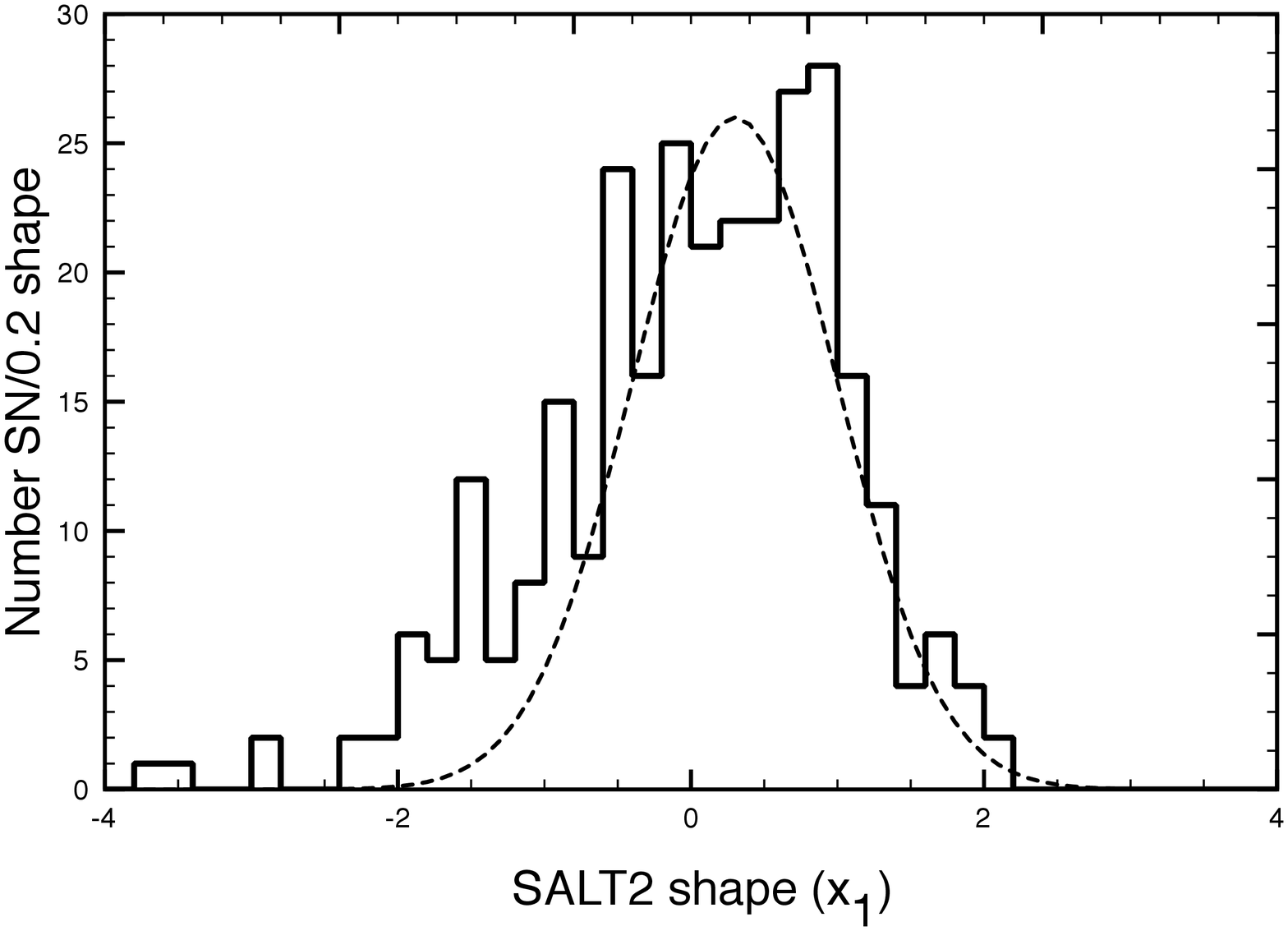}
\caption{The distribution in SALT2 shape parameter ($x_1$) of 296 SDSS SN from our sample of 343 SN that were also determined to be hosted by star-forming galaxies according to the analysis of \citep{La10} is shown as a histogram and the dashed curve is Gaussian of width 1.  The Gaussian is a good description of the core of the distribution but does not describe the tail towards low $x_1$. \label{fig:x1}}
\end{figure}

Unfortunately, we do not have any useful \textit{a priori} model for the $x_1$ distribution.  We can, however, make the somewhat trivial observation that the intrinsic scatter cannot be larger than the width of the distribution.  It is known that passive and star forming galaxies have different distributions in $x_1$ and, if we assume that the intrinsic scatter is the same for both, we can obtain an upper limit for the intrinsic scatter from the distribution that excludes passive galaxies.  The distributions in $x_1$ were shown by \citet{La10} for the SDSS data, and we have used that analysis for the 296 SN this paper has in common with that analysis.  The resulting distribution is shown in Figure \ref{fig:x1} along with a Gaussian of width 1.0.  The Gaussian is not a good fit to the tails of the distribution, but it is a good match to the core and we conclude that an upper limit for the intrinsic scatter in $x_1$ is 1.0, a limit that is greater than the amount required to explain the scatter in the Hubble diagram.

The simulated data exhibits a significant correlation between $m_x$ and $c$, and it is interesting to ask whether that correlation should be expected in the data.  In the simulation, the correlation arises because the SALT2 model defines $B$-band to be the reference point for the color and because the measured data is largely redward of $B$-band.  The assumed independent variation of each measured band results in the correlation between peak magnitude and color.  While the model of color smearing is arbitrarily constructed to reproduce the scatter in the Hubble diagram, there is some evidence to support independent fluctuations in each of the filters.  As pointed out in \citep{K10}, if the redshift is fit from the light curve data, the dispersion of the results around the spectroscopically determined redshifts is larger than can be accounted for by photometric errors alone, but the color smearing model of SNANA gives approximately the right dispersion.  Based on these considerations, we conclude that the correlation between $m_x$ and $c$ is likely to be positive (as seen in the simulations) for the SDSS data.

While we have been able to make some observations on the elements of the intrinsic scatter matrix, our picture is incomplete and rests on assumptions of uncertain validity.  An alternative approach is to marginalize over all possible intrinsic scatter matices.  We implement this marginalization by running SALT2mu repeatedly with randomly generated intrinsic scatter matrices.  We start with a vector randomly generated on the unit sphere subject to the restriction that all components of the vector are positive.  We then multiply the three components of the vector by 0.2, 1.0, and 0.07 respectively and set $\sigma_0$, $\sigma_1$, and $\sigma_c$ to the scaled vector components.  The constants are chosen to be about 50\% larger than the values (see Table \ref{tab:meas}) that were required to explain the intrinsic scatter.  The normalized correlations are chosen randomly between $-1$ and $+1$.  

\begin{figure}
\epsscale{1.0}
\plotone{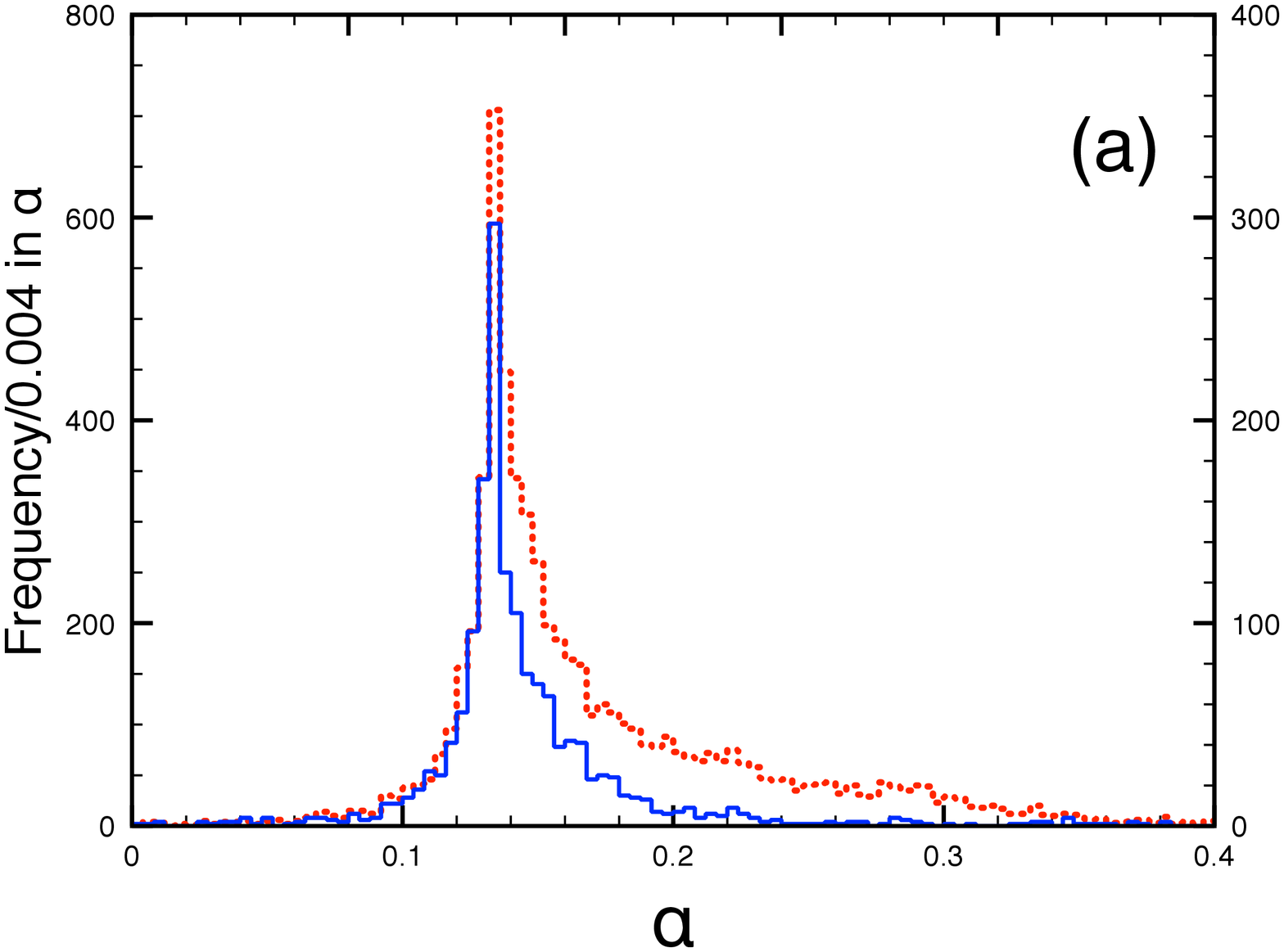}
\plotone{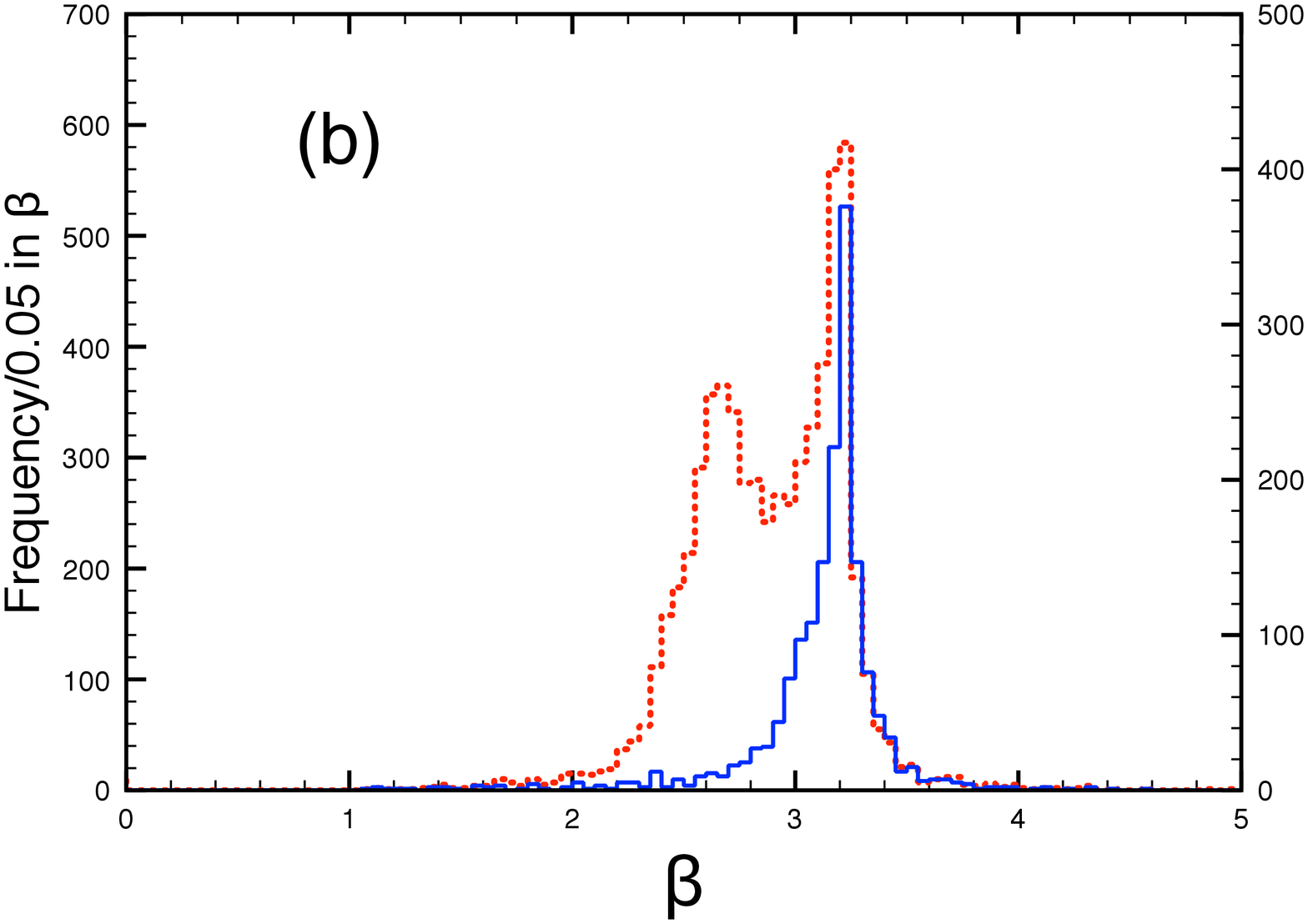}
\caption{The distribution of SALT2mu fit results for $\alpha$ (a) and $\beta$ (b) for a random distribution of intrinsic scatter matrices that yield $\chi^2$ per degree of freedom equal to 1 is shown as the dotted red curve..  The solid blue curve shows the more restricted subset of error matrices that satisfy $\sigma_1<1$, $0.040<\sigma_c<0.082$, and $\xi_{0c}>0$.  The left axis shows the vertical scale for the red curve; the right axis is the scale for the blue curve.  Of the 1650 entries in the restricted subset of intrinsic scatter matrices, there are 32 with $\alpha<0$, 15 with $\alpha>0.4$, 1 with $\beta<0$ and 15 with $\beta>5$, which are not shown in the figures.  The underflow and overflow entries result from cases where the entire distribution can be explained by the assumed intrinsic scatter of the data and the parameters $\alpha$ and $\beta$ become indeterminate, resulting in large values and large estimated errors for the parameters.\label{fig:abmarg}}
\end{figure}

A total of 10000 intrinsic scatter matrices were generated, but the procedure does not guarantee that the matrices are positive.  A total of 6380 of the matrices were found by SALT2mu to be positive definite and were used to determine values for $\alpha$ and $\beta$. Each intrinsic scatter matrix is scaled by an arbitrary factor to achieve a $\chi^2/dof=1$.  The red (dotted) curve in Figure \ref{fig:abmarg} shows the results of the fits for (a) $\alpha$ and (b) $\beta$.  The median value is $\alpha=0.150_{-0.021}^{+0.083}$  and $\beta=2.91_{-0.35}^{+0.30}$, where the error is the range that includes 68\% of the results.  Our intrinsic scatter covariance matrix defines an intrinsic error in 3 dimensions, but the chi-squared is only sensitive to the error in one direction.  The intrinsic errors in the other two directions just broaden the distribution of points in the plane defined by Equation \eqref{eqn:mu}.  If the intrinsic errors are large enough to explain, by themselves, the observed population, the slopes in Equation \eqref{eqn:mu} become undefined and produce the outliers that are observed in Figure \ref{fig:abmarg}.  The data show a secondary peak near $\beta=2.5$, but this should not be interpreted as evidence for 2 populations with different slopes.  The shape of the plot in Figure 5, of course, depends on the distribution of covariance matrices, which we chose to be uniform.  The lower values of $\beta$ result primarily from intrinsic scatter matrices with a small $\sigma_c$.  The fit becomes insensitive to $\sigma_c$ as $\sigma_c\to0$ and the peak at $\beta=2.5$ is an artifact of our choice of distribution for the intrinsic scatter matrices and the insensitivity of the fit parameter $\beta$ in the vicinity of $\sigma_c=0$.

The blue (solid) curve is the result of restricting the intrinsic scatter matrix to a range favored by the arguments in this section.  Specifically we require the intrinsic scatter matrix color element to be within 3 standard deviations of the value determined by the fit to Equation \eqref{eqn:cdist}, namely $0.040<\sigma_c<0.082$.  We also require $\sigma_1<1$ since the intrinsic scatter cannot exceed the measured width of the distribution, and $\xi_{0c}>0$ since we expect a positive correlation when the color smearing includes a component that is uncorrelated between the observed filter bands.  While the exact values of these cuts are somewhat arbitrary, we have chosen a wide range to account for model uncertainties.  Only 1650 of the 6380 intrinsic scatter matrices satisfy these three criteria, but even with the restrictions a wide range of values are found.  However, the median values and the ranges that contains 68\% of the data are $\alpha=0.135_{-0.013}^{+0.031}$ and $\beta=3.19_{-0.22}^{+0.10}$.  In addition to the range of values that result from different assumptions about the intrinsic scatter matrix, the statistical errors reported by SALT2mu are $\sigma_\alpha=0.011$ and  $\sigma_\beta=0.10$.  Adding these errors in quadrature, the final result is $\alpha=0.135_{-0.017}^{+0.033}$ and $\beta=3.26_{-0.24}^{+0.14}$.  For each entry in Figure \ref{fig:abmarg}, we compute the statistical probability that the entry is consistent with  $\beta \geq 4.1$ using the computed statistical error and assuming a normal distribution.  The uncertainty in the intrinsic scatter matrix is computed by averaging the statistical probability for all the entries in Figure \ref{fig:abmarg}. From this calculation, we conclude that the probability that our result is consistent with $\beta \geq 4.1$ is 2\%.  The probability is higher than would be naively estimated from the value $\beta=3.26_{-0.24}^{+0.14}$ because our calculation includes the effect of the non-Gaussian tail of the distribution shown in Figure \ref{fig:abmarg}b.

\section{Redshift Dependence of $\alpha$ and $\beta$ in the SDSS data}\label{sec:red}

\begin{figure}
\epsscale{1.0}
\plotone{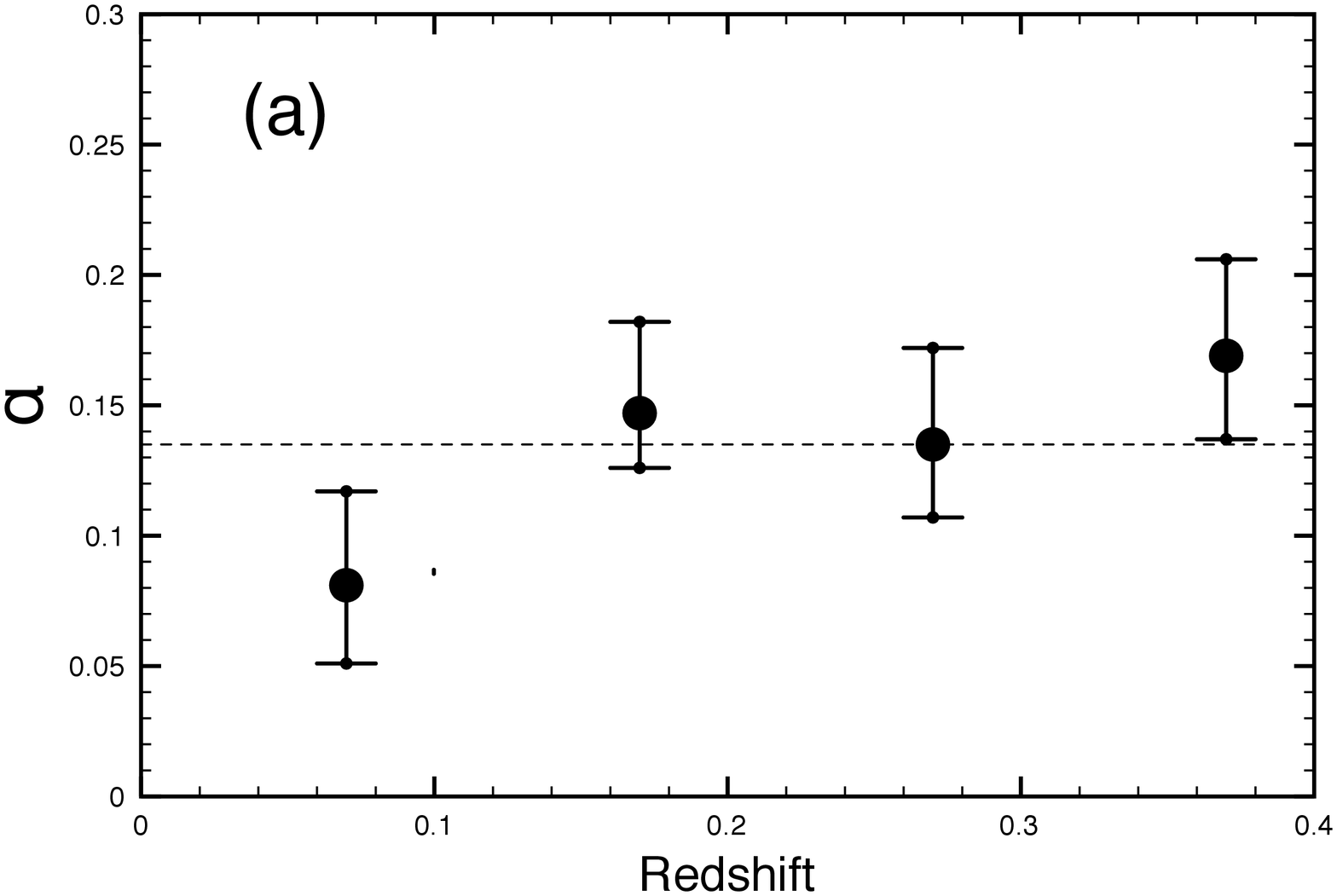}
\plotone{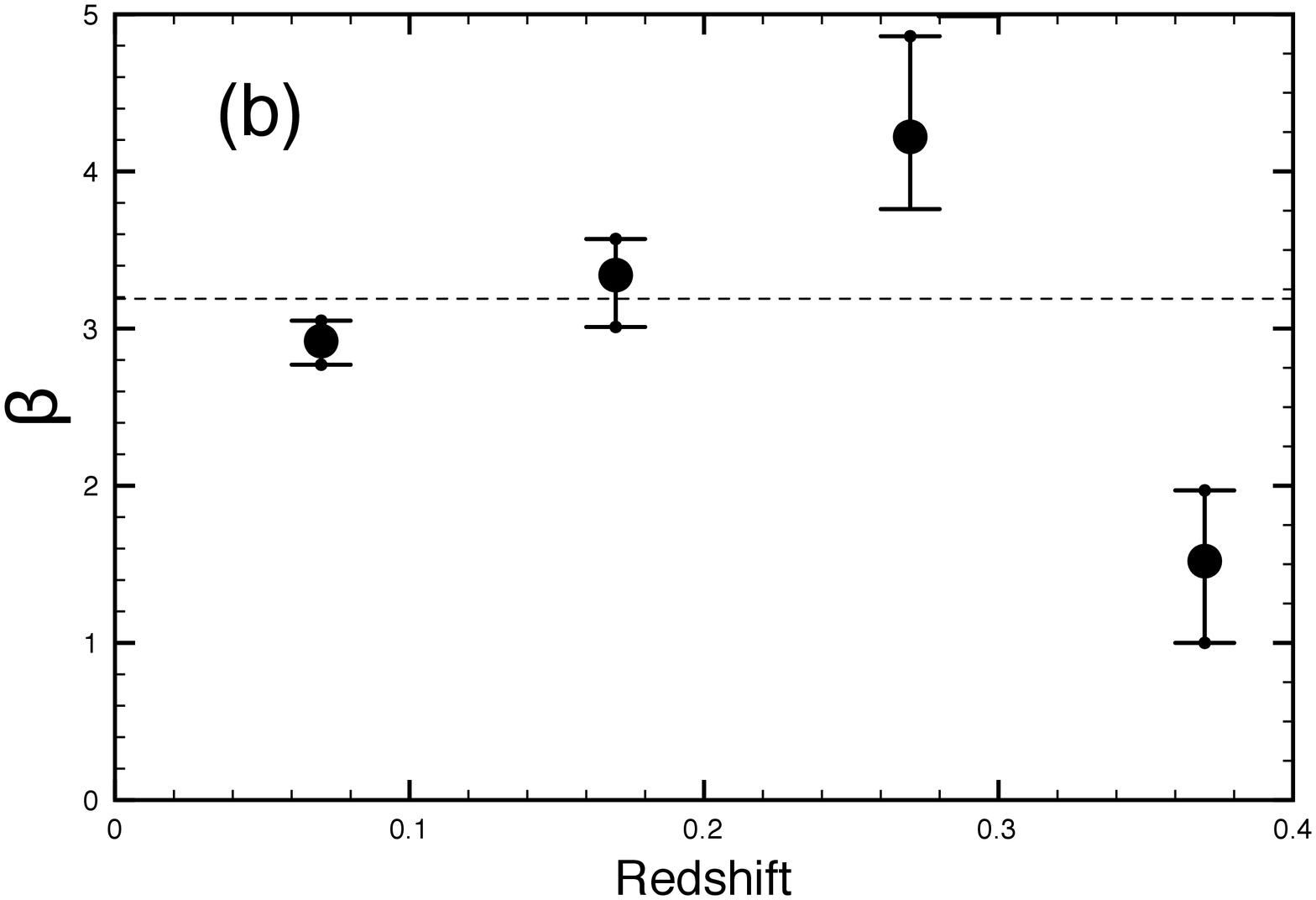}
\caption{The SALT2mu fit results for $\alpha$ (a) and $\beta$ (b) in 4 redshift bins using the same a distribution of intrinsic scatter matrices that yield chi-squared per degree of freedom equal to 1 and satisfy $\sigma_1<1$, $0.040<\sigma_c<0.082$, and $\xi_{0c}>0$.  The results for the full sample are indicated by the dashed lines.\label{fig:abred}}
\end{figure}

The parameters $\alpha$ and $\beta$ are usually assumed to be constants, independent of redshift, but the possibility of a redshift dependence has been considered previously by, for example, \citet{K09b}.  We might observe a redshift dependence of the fits for $\alpha$ and $\beta$ for a number of reasons:
\begin{itemize}
\item{The properties of the SN, themselves, may evolve with redshift.}
\item{The model of Equation \eqref{eqn:stmag} may not be an exact model and the resulting fit parameters may be in tension between different SN populations.  If the population is a function of redshift, the amount of tension may vary, resulting in different results for $\alpha$ and $\beta$.}
\item{Our SN selection has biases, particularly at higher redshift where we tend to select the brighter SN.}
\end{itemize}
While we could directly fit for $\alpha(z)$ and $\beta(z)$, the likelihood that selection effects will play a significant role suggests that a better approach might be to split the SDSS sample into 4 redshift bins.  The results of the fits are shown in Figure \ref{fig:abred}.  We have used the the same ensemble of intrinsic scatter matrices with $\sigma_1<1$, $0.040<\sigma_c<0.082$, and $\xi_{0c}>0$ that was used to construct Figure \ref{fig:abmarg}.  The errors shown are the combined statistical and systematic error that arise from the uncertainty in the intrinsic scatter matrix.  The results shown for $\alpha$  in Figure \ref{fig:abred}(a) are consistent with the sample average although there is an indication that the value of $\alpha$ may be rising as a function of redshift.  Figure \ref{fig:abred}(b) show a value of $\beta$ that is $2.1\sigma$ lower in the lowest redshift bin, $2.2\sigma$ higher in the range $0.22<z<0.32$, and $3.7\sigma$ lower in the highest redshift bin than the result for the complete sample.  The accuracy in the determination of $\beta$ as a function of redshift varies not only because of the statistics but because highly reddened SN are too dim to meet our selection criteria at high redshift.  The number of SN in each redshift bin and the number of highly reddened ($c>0.3$)  SN are:  36 and 4 ($0.02<z<0.12$), 149 and 2 ($0.12<z<0.22$), 116 and 0 ($0.22<z<0.32$), and 41 and 0 ($0.32<z<0.42$), respectively.  The reduced range of $c$ at the higher redshifts reduces the accuracy of the fit and makes the fit considerably more sensitive to the amount of intrinsic scatter in the data. 
The highest-redshift sample consists of 41 SN with $-0.319<c<0.126$ that show very little correlation between the distance modulus and $c$, resulting in a low value of $\beta$.  The range of observed values of $c$ is most likely primarily due to measurement error and intrinsic scatter resulting in the weak correlation between distance modulus and observed $c$.  In principle, our formalism should recover the correct value of $\beta$ when the correct measurement errors and intrinsic scattering are included, but selection effects have significantly distorted the natural spread.  In addition to the restricted color range, the absolute magnitudes of SN with $0.32<z<0.42$ are $0.069\pm0.042$ magnitudes brighter on average, assuming standard cosmological parameters ($\Omega_\Lambda=0.7$ and $w=-1$).  As a consequence, our fit to the last redshift bin should not be considered to be a good representation of what would be obtained with an unbiased sample.  A more robust result would be possible with an improved simulation of the selection effects and augmenting the spectroscopically confirmed sample with those identified photometrically \citep{Sa11} to obtain a more complete sample.
Selection effects should be much less important for the midrange ($0.22<z<0.32$), where is $\beta$ larger than the sample average by $2.2\sigma$.  While the lack of highly reddened SN ($c>0.3$) degrades our ability to determine $\beta$ in this redshift region, we can see if their absence biases $\beta$ to higher values by excluding SN with $c>0.3$ in the low redshift bins.  The $\beta$ values for the two lowest redshift regions change by less than 0.1 when the highly reddened SN are removed from the sample.

\section{Discussion}

The histograms in Figure \ref{fig:abmarg} show the same trend that was observed in Table \ref{tab:meas}:  when the intrinsic scatter in color is large, the best fit value of $\beta$ will increase while $\alpha$ decreases slightly.  The fact that our value for $\beta$ is the highest value of those reported in \S\ref{sec:SDSS} depends on our incorporation of a large intrinsic scatter in supernova color.  We have not made any correction for the selection bias of our sample based on the results of the simulation, which is subject to the same selection criteria, and shows no significant bias in the determination of $\alpha$ and $\beta$.

The error on our result is dominated by an uncertainty in the intrinsic scatter matrix, but there are  other potential problems that have not been taken into account.  Certainly, our assumption that the intrinsic scatter is a constant, independent of any SN properties, is suspect given our lack of knowledge of the processes involved.  In addition, our model assumes that the data follow the relationship of Equation \eqref{eqn:stmag} with a fixed value of $\beta$.  However, we know that the value of $R_V$ varies significantly in our own galaxy \citep{Sc10}, and we should expect to see at least some variation in SN host galaxies.  Variations in $\beta$ produce effects similar to the intrinsic scatter in color, but the effect can be magnified in a magnitude limited sample like SDSS where there are few highly reddened SN, which have higher weight in determining $\beta$.  Variations in $\beta$ could explain the small differences seen between the values of $\beta$ in Figure \ref{fig:abred} in the two lowest redshift bins.

We have seen that neglecting intrinsic scatter in the SALT2 $c$ parameter results in a value of $\beta$ that is biased low.  The amount of bias depends on the amount of intrinsic scatter relative to the range in the variable $c$.  Since the range in $c$ decreases as the redshift increases, neglecting or underestimating the amount of intrinsic scatter in $c$ would be expected to result in a decreasing value of $\beta$ as a function of redshift.  The fits to the SDSS data as a function of redshift include intrinsic scatter in color so this effect should not be the explanation for the SDSS data---unless the scatter was underestimated.  However, an overestimate of the intrinsic scatter in $m_x$ leads to similar biases and is more likely explanation for the lower value of $\beta$ that is observed in the highest redshift bin.

A more problematic type of variation in $\beta$ could arise from the color smearing.  We have argued that color variations of SN must have some component of host galaxy extinction and also some intrinsic color variation, which could be due to variations in the explosion process or local environment or both.  We have modeled the intrinsic color variation as color smearing with an effective value $\beta=0$, but it is possible that each mechanism that contributes to the observed color requires a different correction to the distance modulus.  Our model can only produce an effective value of $\beta$ that could be a function of color, redshift, host galaxy type or other parameters and could, therefore, differ depending on the characteristics of the particular SN sample.

Although we have not included correlations with galaxy types or spectral line features, the introduction of additional dependent variables into Equation \eqref{eqn:stmag} is straightforward.  In fact, having a more complete description of the supernova explosion will not only provide a smaller dispersion for the numerator in Equation \eqref{eqn:s2mchi2}, it will decrease the size of the intrinsic scatter that is required and decrease the importance of the uncertainty as to its form.

Recently a number of papers \citep{Ki11,La11,Ma11} have explored the technique of determining the intrinsic scatter from the data.    The approach of \citep{Ma11} is similar to ours in that it included our more general form for the intrinsic error matrix but assumed that the off-diagonal terms involving $M_B$ were zero.  We have not considered including the intrinsic scattering error as a fit parameter partly for simplicity and partly because the underlying physics of the intrinsic scatter is still uncertain.  However, the more rigorous statistical approach described by these authors is a possible direction for future work.

We argue from the SDSS SN photometric data that there is likely some intrinsic scatter in SN color that is responsible for some of the scatter in the Hubble diagram. A similar conclusion is reached by \citet{Fo11}, based on much more detailed information involving differences in the line velocities measured from the SN spectra, which are modeled as arising from differences in viewing angle.  While this type of detailed spectral information may not be available for all future SN surveys, an understanding of the underlying physical processes will be important for the most accurate treatment of the data.

\section{Summary}

We have described a new formalism to fit the parameters $\alpha$ and $\beta$ that are used to determine the standard magnitudes of Type Ia supernovae.  The determination of these parameters is accomplished in a way that is independent of cosmology by introducing a few nuisance parameters to accommodate any reasonable cosmology.  The formalism also introduces a more general form for the intrinsic scatter which introduces a large uncertainty into the fit if the parameters describing the intrinsic scattering are not known accurately.  We have shown by simulation that the mathematical formalism is self-consistent, but have pointed out ways in which the model may fail to be a complete description of SN.

We find that the SDSS data, when fit to the form of Equation \eqref{eqn:stmag} are described by $\alpha=0.135_{-0.017}^{+0.033}$ and $\beta=3.19_{-0.24}^{+0.14}$ by marginalizing over the uncertainty in the intrinsic scatter matrix.  Our result relies on our conclusion that SNe are subject to a substantial color smearing in addition to reddening from host galaxy dust, as indicated by our fit to the color distribution where we determined that $\sigma_c=0.055\pm0.007$.  The uncertainty in the parameters of the intrinsic scatter matrix results in a much larger error than is obtained if the parameters are assumed to be known.  However, even with the larger value of $\beta$ and the larger error, we find that SDSS data differ at the 98\% confidence level from $\beta=4.1$, the value expected for extinction by dust in the Milky Way. 

\section{Acknowledgements}

The SDSS-II SN survey was managed by the Astrophysical Research Consortium for the Participating Institutions. The Participating Institutions were the American Museum of Natural History, Astrophysical Institute Potsdam, University of Basel, Cambridge University, Case Western Reserve University, University of Chicago, Drexel University, Fermilab, the Institute for Advanced Study, the Japan Participation Group, Johns Hopkins University, the Joint Institute for Nuclear Astrophysics, the Kavli Institute for Particle Astrophysics and Cosmology, the Korean Scientist Group, the Chinese Academy of Sciences (LAMOST), Los Alamos National Laboratory, the Max-Planck-Institute for Astronomy (MPA), the Max-Planck-Institute for Astrophysics (MPiA), New Mexico State University, Ohio State University, University of Pittsburgh, University of Portsmouth, Princeton University, the United States Naval Observatory, and the University of Washington.

This work is based in part on observations made at the following telescopes. The Hobby- Eberly Telescope (HET) is a joint project of the University of Texas at Austin, the Pennsylvania State University, Stanford University, Ludwig-Maximillians-Universit\"at M\"unchen, and Georg-August-Universit\"at G\"ottingen. The HET is named in honor of its principal benefactors, William P. Hobby and Robert E. Eberly. The Marcario Low-Resolution Spectrograph is named for Mike Marcario of High Lonesome Optics, who fabricated several optical elements for the instrument but died before its completion; it is a joint project of the Hobby-Eberly Telescope partnership and the Instituto de Astronom\'ia de la Universidad Nacional Aut\'onoma de M\'exico. The Apache Point Observatory 3.5 m telescope is owned and operated by the Astrophysical Research Consortium. We thank the observatory director, Suzanne Hawley, and former site manager, Bruce Gillespie, for their support of this project. The Subaru Telescope is operated by the National Astronomical Observatory of Japan. The William Herschel Telescope (WHT) is operated by the Isaac Newton Group, the Nordic Optical Telescope (NOT) is operated jointly by Denmark, Finland, Iceland, Norway, and Sweden, and the Telescopio Nazionale Galileo (TNG) is operated by the Fundaci\'on Galileo Galilei of the Italian INAF (Istituto Nazionale di Astrofisica) all on the island of La Palma in the Spanish Observatorio del Roque de los Muchachos of the Instituto de Astrof\'isica de Canarias. Observations at the ESO New Technology Telescope at La Silla Observatory were made under programme IDs 77.A-0437, 78.A-0325, and 79.A-0715. Kitt Peak National Observatory, National Optical Astronomy Observatories (NOAO), is operated by the Association of Universities for Research in Astronomy, Inc. (AURA) under cooperative agreement with the NSF. The South African Large Telescope (SALT) of the South African Astronomical Observatory is operated by a partnership between the National Research Foundation of South Africa, Nicolaus Copernicus Astronomical Center of the Polish Academy of Sciences, the Hobby-Eberly Telescope Board, Rutgers University, Georg- August-Universit\"at G\"ottingen, University of Wisconsin-Madison, University of Canterbury, University of North Carolina-Chapel Hill, Dartmouth College, Carnegie Mellon University, and the United Kingdom SALT consortium. The WIYN Observatory is a joint facility of the University of Wisconsin- Madison, Indiana University, Yale University, and NOAO.  The W.M. Keck Observatory is operated as a scientific partnership among the California Institute of Technology, the University of California, and the National Aeronautics and Space Administration. The Observatory was made possible by the generous financial support of the W. M. Keck Foundation.

This work was supported in part by the U.S. Department of Energy under contract number DE-AC0276SF00515.

\end{document}